# Efficient third-harmonic generation in high-Q tantalum pentoxide microcavities

XINKE XING,[1,2] YIXUAN YANG,[2,3] XINZHI ZHENG,[4,5] RENHONG GAO,[5,10] LI DENG,[5,11] JINTIAN LIN,[2,3,*] YA CHENG,[4,5,6,7,8,9,12]

[1] *School of Physics Science and Engineering, Tongji University, Shanghai 200092, China*
[2] *State Key Laboratory of Ultra-Intense Laser Science and Technology, Shanghai Institute of Optics and Fine Mechanics, Chinese Academy of Sciences, Shanghai 201800, China*
[3] *Center of Materials Science and Optoelectronics Engineering, University of Chinese Academy of Sciences, Beijing 100049, China*
[4] *State Key Laboratory of Precision Spectroscopy, East China Normal University, Shanghai 200062, China*
[5] *The Extreme Optoelectromechanics Laboratory (XXL), School of Physics, East China Normal University, Shanghai 200241, China*
[6] *Shanghai Research Center for Quantum Sciences, Shanghai 201315, China*
[7] *Hefei National Laboratory, Hefei 230088, China*
[8] *Collaborative Innovation Center of Extreme Optics, Shanxi University, Taiyuan 030006, China*
[9] *Collaborative Innovation Center of Light Manipulations and Applications, Shandong Normal University, Jinan 250358, China*
[10] *E-mail: rhgao@phy.ecnu.edu.cn*
[11] *E-mail: ldeng@phy.ecnu.edu.cn*
[12] *E-mail: ya.cheng@siom.ac.cn*
* *E-mail: jintianlin@siom.ac.cn*

**Nonlinear harmonic generation in microcavities offers a compelling route toward energy efficient on-chip short-wavelength coherent light sources, enabled by tight optical confinement within compact mode volumes and flexible dispersion engineering. Tantalum pentoxide ($Ta_2O_5$) photonic platforms are particularly promising for realizing such nonlinear processes: their broad transparency window (280–8000 nm) supports broadband operation, their large bandgap suppresses two-photon absorption, and their thick-film deposition capability enables dispersion management—features that can collectively enhance the performance of nonlinear light sources. Here, we demonstrate, to the best of our knowledge, the first efficient third-harmonic generation (THG) in high-Q $Ta_2O_5$ microcavities. These microcavities feature smooth surfaces and high loaded Q factors up to $2.67\times10^6$, thereby facilitating nonlinear interactions. Through dispersion engineering, efficient THG is realized in both suspended microdisks and monolithic microrings under telecom-band continuous-wave laser pumping, as confirmed by optical spectra, visible-light microscopy, and the near-cubic pump-power dependence of the conversion efficiency. This work lays a foundation for chip-scale *f*–3*f* self-referencing of frequency combs and next-generation integrated nonlinear photonics.**

Integrated nonlinear photonics which significantly extend spectral coverage of both classical and quantum light sources, has emerged as a key enabling technology for constructing compact, energy-efficient, and scalable optical systems [1-6]. In particular, the generation of short-wavelength coherent light in chip is highly desirable for applications ranging from precision metrology, optical manufacturing, and spectroscopy to optical clocks and quantum photonics [7-11]. Whispering-gallery-mode (WGM) optical microcavities offer a compelling platform for this purpose because they confine light with a long period of time within a small mode volume via continuous total internal reflection along the circular boundary, thereby dramatically enhancing light–matter interactions and reducing the pump power required for nonlinear frequency conversion [12]. Moreover, flexible dispersion engineering in microcavities allows phase matching and broadband operation to be tailored within compact footprints, making them attractive for efficient nonlinear harmonic generation [13-19].

Among nonlinear optical frequency conversion processes, third-harmonic generation (THG) is of particular interest because it provides a direct route to visible and ultraviolet coherent light from near-infrared pump sources. However, efficient THG in integrated platforms remains challenging: it requires simultaneously high optical confinement, low propagation loss (corresponding to high Q factors), broad transparency window, a wide bandgap, and precise dispersion control. The choice of material platform is therefore critical. Tantalum pentoxide ($Ta_2O_5$) has recently attracted attention as a promising nonlinear photonic platform [20-28]. Its broad transparency window (280–8000 nm) supports wideband operation, its moderate refractive index (2.05) allows tight optical confinement, its large bandgap (3.8 eV) enables the avoidance of two-photon absorption in short wavebands, high nonlinear coefficient (~$7.2\times10^{-19}$ $m^2/W$) facilitates nonlinear conversion, and its thick-film deposition capability with low residual-stress enables effective dispersion management.

These features make $Ta_2O_5$ particularly suitable for enhancing the performance of integrated nonlinear light sources. Nevertheless, efficient THG in high-Q $Ta_2O_5$ microcavities has remained unexplored.

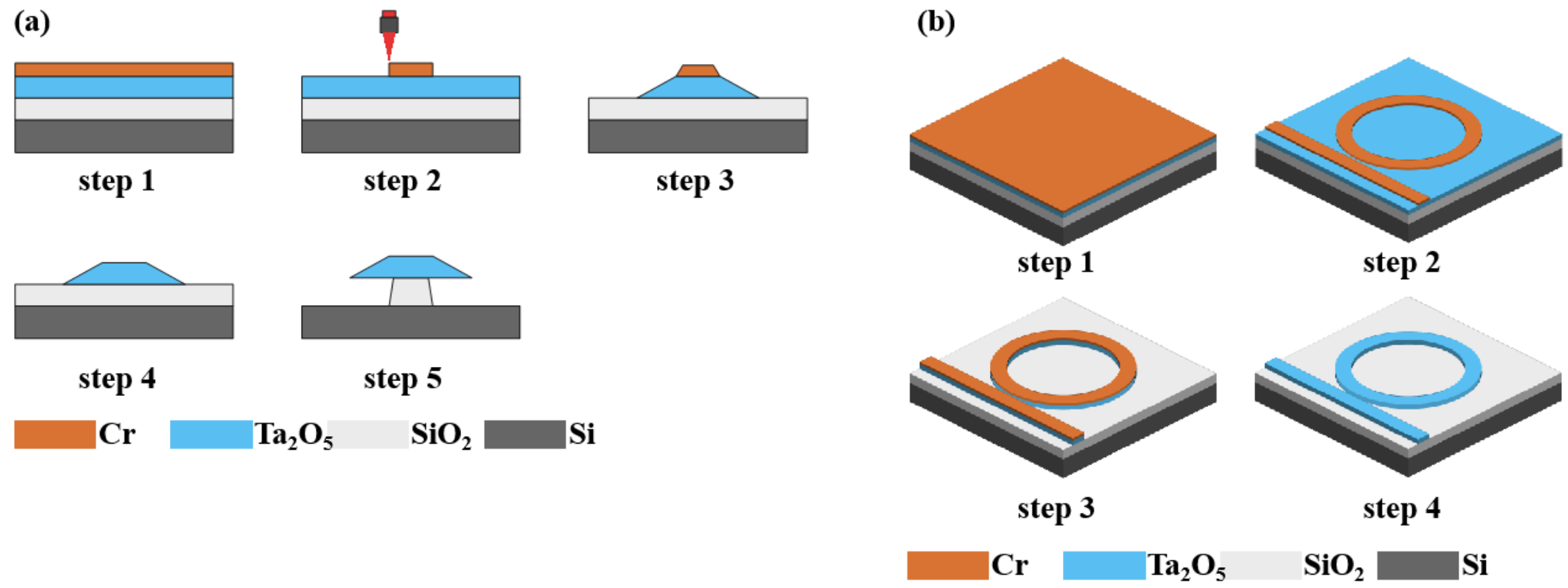

Fig. 1 Schematic of the fabrication flow for (a) microdisks and (b) microrings. The process comprises several steps: (1) successively depositing a $Ta_2O_5$ layer and a chromium (Cr) layer on a thermally oxidized silicon wafer; (2) patterning the Cr film into a microdisk or a microring side-coupled to a bus waveguide by femtosecond laser ablation to form a hard mask; (3) transferring the Cr mask pattern into the $Ta_2O_5$ film by chemo-mechanical polishing (CMP); (4) removing the Cr mask by wet chemical etching and improving the surface quality via a second CMP step. For microdisk fabrication, the $SiO_2$ layer underneath the microdisks is undercut by wet etching to form suspended microdisks (step 5).

In this work, to the best of our knowledge, the first efficient THG is demonstrated in high-Q $Ta_2O_5$ microcavities. The microcavities feature smooth surfaces and loaded Q factors up to $2.67\times10^6$, which facilitate strong nonlinear interactions. By engineering the dispersion, we realize THG in both suspended microdisks and monolithic microrings. The generated third harmonic is confirmed through optical spectra, visible-light microscopy, and measurements of the pump-power dependence of the conversion efficiency. We further characterize the temperature dependence of the THG efficiency, which exhibits a Lorentzian line-shape with a full width at half maximum (FWHM) of 8.47°C. These results establish high-Q $Ta_2O_5$ microcavities as a viable platform for efficient integrated nonlinear frequency conversion.

## 2. Fabrication and Structure Characterization

High-Q $Ta_2O_5$ microresonators (including microdisks and microrings) were fabricated using the femtosecond laser photolithography assisted chemo-mechanical etching (PLACE) technique [20-22], as shown in Fig. 1. For the microdisks, a 900-nm-thick amorphous $Ta_2O_5$ film was first deposited by magnetron sputtering onto a commercially available 4.7-µm-thick thermally grown $SiO_2$ layer on a Si wafer, forming a $Ta_2O_5$-on-insulator wafer. An 800-nm-thick chromium (Cr) layer was then deposited on the $Ta_2O_5$ surface by magnetron sputtering. Next, spatially selective femtosecond-pulse laser ablation patterned the Cr layer into microdisks with a spatial resolution of 200 nm. The patterned Cr structures served as hard masks for the subsequent etching owing to their high hardness. The Cr mask pattern was transferred into the $Ta_2O_5$ film by chemo-mechanical polishing (CMP), which selectively removed the exposed $Ta_2O_5$ regions while leaving the Cr-masked regions intact. Subsequently, the residual Cr masks were removed with a Cr etchant. A second CMP step was performed to further improve the surface smoothness of the $Ta_2O_5$ microdisk and reduce residual surface defects, yielding an ultrasmooth microdisk surface. Finally, the $SiO_2$ layer beneath the microdisk was etched to form a small supporting pedestal, resulting in a suspended microdisk. For the microring resonators, they were fabricated using the same process, except that the deposited amorphous $Ta_2O_5$ film was 600 nm thick and the femtosecond-laser-patterned Cr mask defined a microring side-coupled with a bus waveguide.

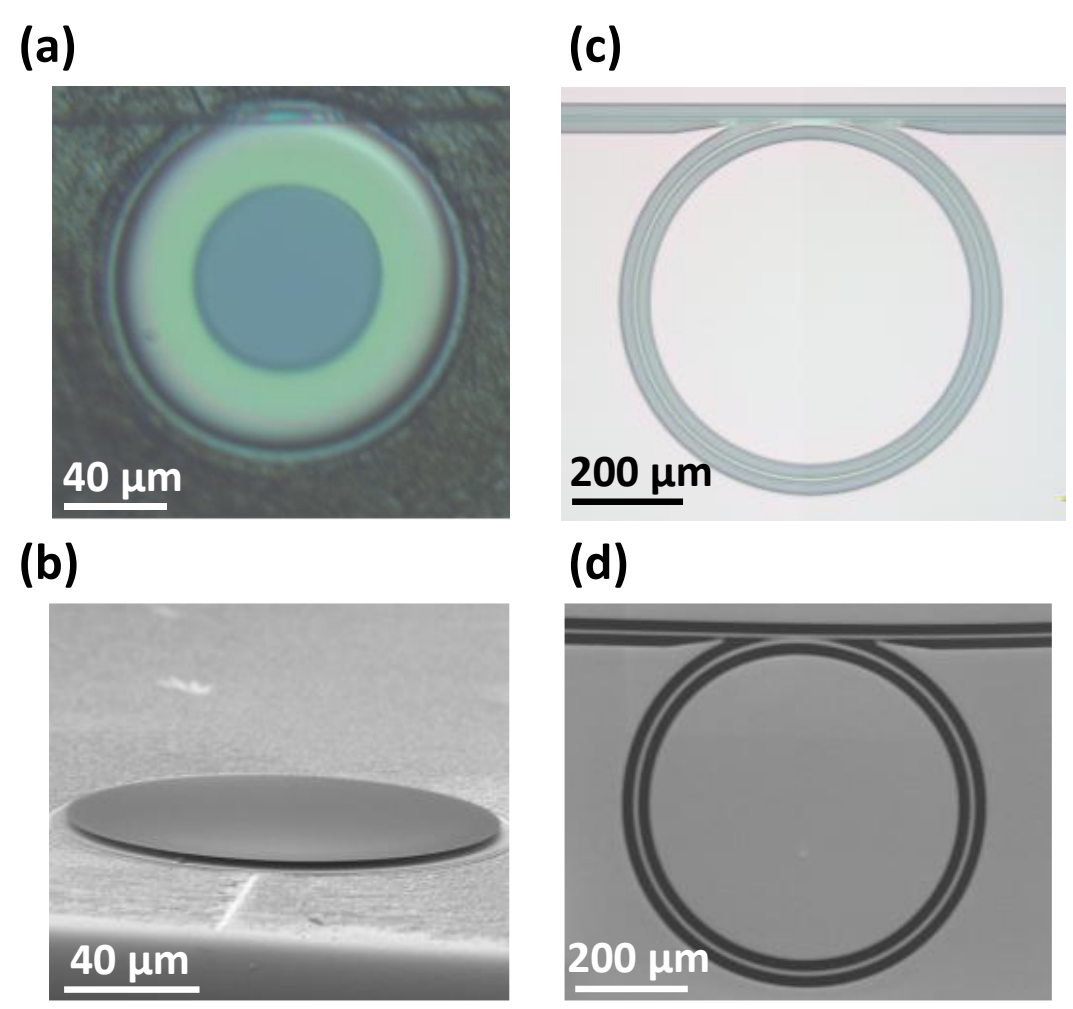

Fig. 2. Morphologies of the microcavities. (a) Optical micrograph and (b) scanning-electron-microscope (SEM) image of the microdisk. (c) Optical micrograph and (d) SEM image of the microring.

The fabricated suspended microdisk has a radius of 60 µm and is supported by a small pedestal, as shown in its optical micrograph and scanning electron microscope (SEM) image (Figs. 2(a) and 2(b)). The microring has a radius of 300 µm and a top width of 4 μm, while the bus waveguide has a top width of 2 μm and is separated from the microring by 6.5 μm, as shown in Figs. 2(c) and 2(d).

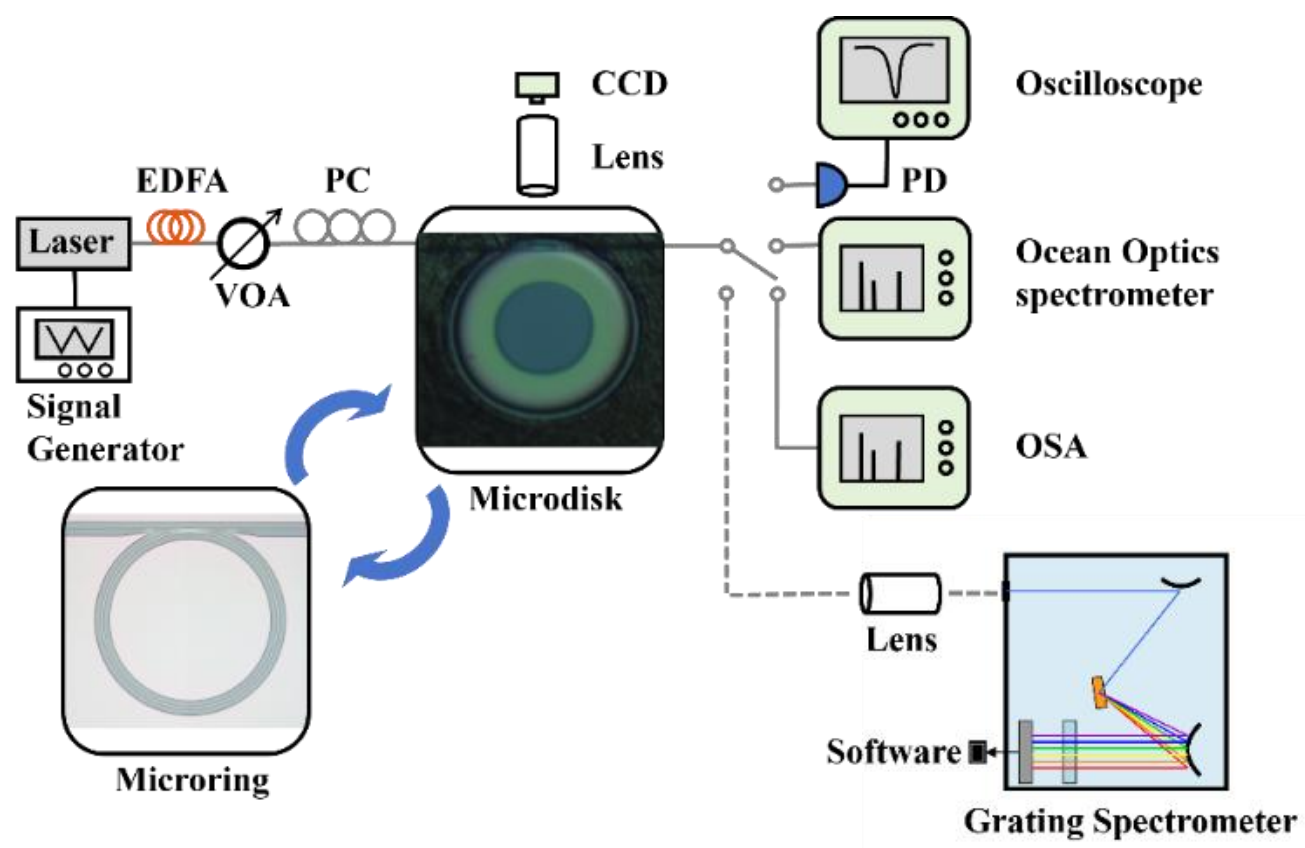

Fig. 3. Experimental setup for WGM characterization and spectral measurement. VOA: variable optical attenuator. PC: polarization controller. CCD: charge-coupled device. PD: photodetector. OSA: optical spectrum analyzer.

Figure 3 shows the experimental setup for THG in both the microdisk and the microring. A C-band tunable laser (TLB-6728, New Focus Inc.), amplified by an erbium-doped fiber amplifier (EDFA) and with its power adjusted by a variable optical attenuator (VOA), was used as the continuous-wave (CW) pump. An in-line fiber polarization controller (PC) was used to set the pump to transverse-electric (TE) polarization. For microdisk coupling, light was coupled into and out of the microdisk through a tapered fiber with a 2-μm beam waist placed in contact with the top surface of the microdisk. An optical microscope imaging system, consisting of a microscope objective with a numerical aperture (NA) of 0.28 and a visible–infrared charge-coupled device (CCD), was positioned directly above the microdisk to enable real-time monitoring of the coupling. The output was delivered through the tapered fiber taper to an optical spectrum analyzer (OSA; Yokogawa AQ6370D; detection range: 600–1700 nm) and an Ocean Optics spectrometer (USB 2000; detection range: 300–1000 nm) for spectral analysis of the pump and THG signals, respectively. Nonlinear emission from the microcavity surface was captured by the optical microscope imaging system. To characterize the Q factors of the microcavity, the output was detected by a photodetector (PD; 1611-FC, New Focus Inc.) connected directly to an oscilloscope (MDO3104, Tektronix Inc.). A signal generator linearly scanned the tunable laser wavelength across the resonance wavelength of a WGM, and the transmission spectra were recorded using the oscilloscope.

For microring coupling, the experimental setup was similar to that for the microdisk, except that a lensed fiber replaced the tapered fiber for coupling to the bus waveguide. The pump light and the Q factor of the pump mode were measured using the same oscilloscope-based scheme. To avoid the significant loss associated with fiber transmission, the THG signal output from the waveguide facet was imaged onto a grating spectrometer (Kymera 328i-B2, Andor Technology Ltd.) using an objective lens for spectral analysis.

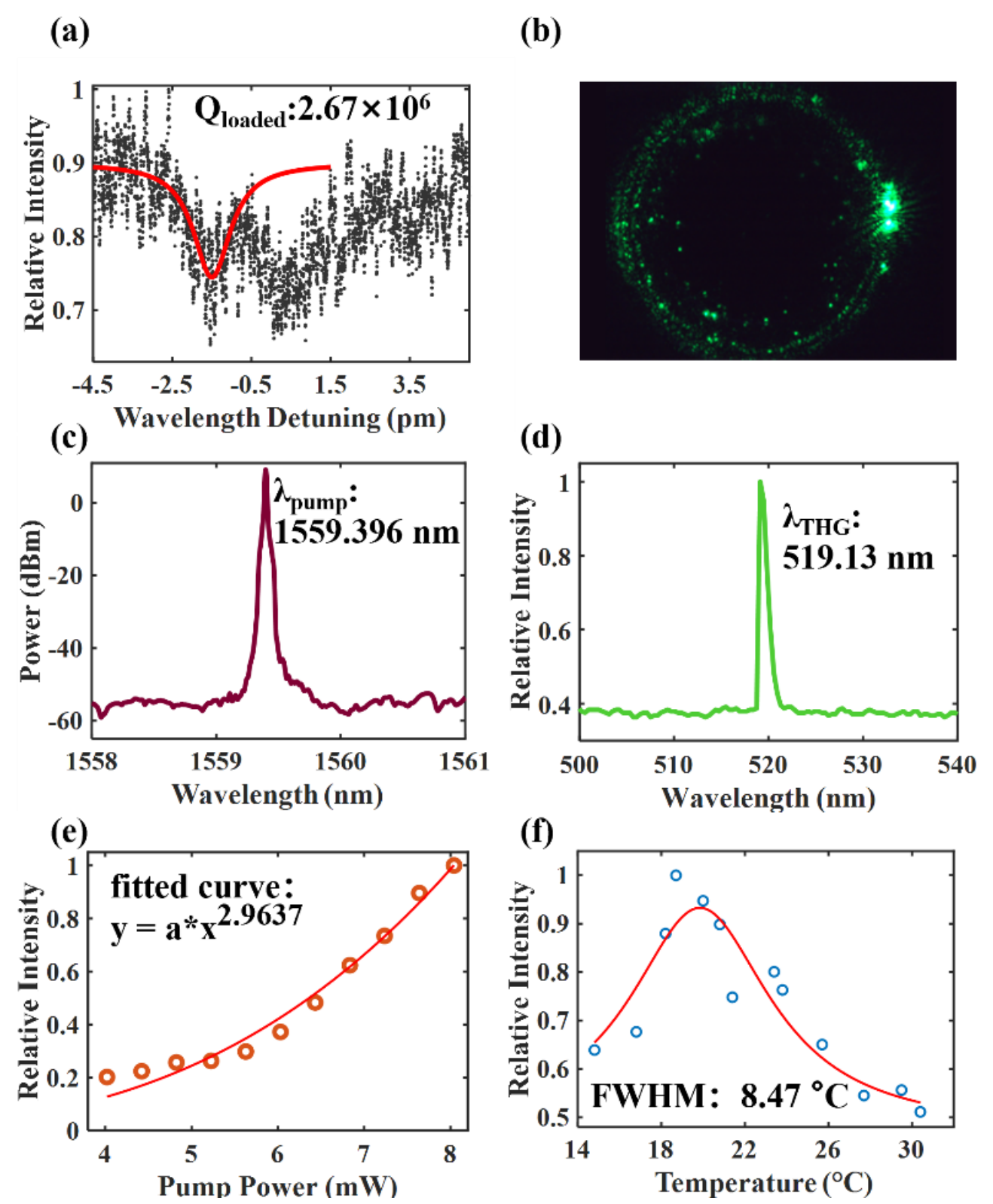

Fig. 4. (a) Loaded Q factor of the pump mode of microdisk.. (b) Optical micrograph of the scattered green light from the microdisk. (c) Optical spectra of the pump light and (d) THG signal. (e) Measured THG power as a function of the pump input power. (f) THG intensity response on temperature optimization.

The loaded Q of the pump mode of the microdisk cavity was measured to be $2.67\times10^6$ (Fig. 4(a)). When the pump mode at 1559.40 nm was excited by a pump power of 12.4 mW, bright green light emission from the microdisk surface was observed in the optical microscope imaging system, as shown in Fig. 4(b), exhibiting a characteristic high-order spatial WGM resonance. A nonlinear signal was detected at 519.13 nm, which is exactly one-third of the pump wavelength, as evidenced by the spectra of the pump and nonlinear signals shown in Figs. 4(c) and 4(d), respectively. According to the energy conservation, this nonlinear process is attributed to THG. The dependence of the nonlinear signal intensity on the pump power was experimentally measured for pump powers ranging from 4 to 8 mW. A logarithmic fit to the data yields a THG dependence on the pump power ($x$) of $x^{2.9637}$, deviating by only 1.21% from the expected cubic depnedence, which further confirms the THG nature of the process. Because the pump and THG modes of the microdisk belonging to different mode families, exhibit different thermal tuning characteristics, the THG efficiency is governed by and can be optimized through temperature variation. At an on-chip pump power of 12.4 mW, the tunability of the THG efficiency was investigated using a temperature controller. The THG intensity exhibits a Lorentzian line shape as a function of temperature [12], and the fitted curve has a full width at half maximum (FWHM) of 8.47 °C, as shown in Fig. 4(f).

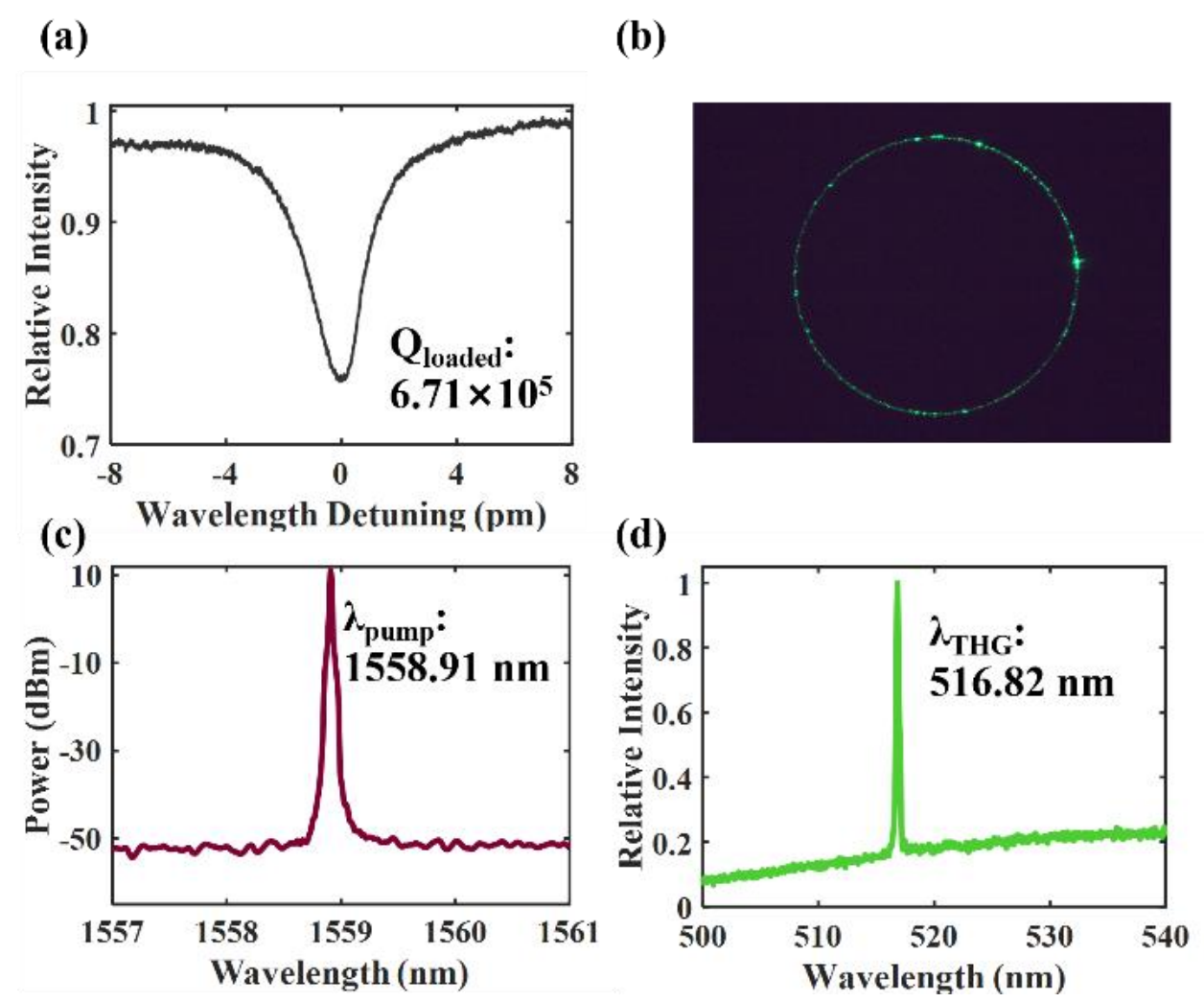

Fig. 5. (a) Loaded Q factor of the pump mode of microring. (b) Optical micrograph of the scattered light from the microring. (c) Optical spectra of the pump light and (d) THG signal.

For THG in the microring, the end-face coupling loss between the waveguide facet and the lensed fiber was measured to 5.718 dB/facet. The loaded Q factor of the pump mode in the microring was also measured at low power and was found to be $6.71 \times 10^5$. The THG signal was observed at 516.82 nm when the pump wavelength was set to 1558.91 nm and the on-chip pump power was 47.08 mW, as evidenced with the spectra of the pump and THG signals shown in Figs. 5 (c) and (d), respectively. Visible light emission was also recorded, showing a green light profile along the ring periphery, as evidenced in Fig. 5(b). Therefore, efficient THG was demonstrated in both the high-Q microdisk and microring, under CW laser pumping.

In summary, we have fabricated high-Q $Ta_2O_5$ microcavities for efficient THG. THG signals are observed under telecom-band pumping and are confirmed by optical spectral analysis, optical microscope imaging, a near-cubic power dependence, and the temperature-tuning response. To the best of our knowledge, this is the first demonstration of integrated THG in $Ta_2O_5$ microresonators. This work provides a foundation for chip-scale $f$–$3f$ self-referencing of frequency combs and for next-generation integrated nonlinear photonics. More broadly, it highlights the potential of dispersion-engineered $Ta_2O_5$ microcavities for compact short-wavelength coherent light sources and for a wide range of on-chip nonlinear optical functionalities.

## References

1. M. G. Vazimali and S. Fathpour, Adv. Photon. **4**(3), 034001 (2022).
2. Z. Xie, F. Bo, J. Lin, *et al.*, Adv. Phys. X **9**(1), 2322739 (2024).
3. K. Zhang, Y. Chen, C. Xu, *et al.*, Chin. Opt. Lett. **23**(4), 041902 (2025).
4. B. Xu, X. Xiong, R. Niu, *et al.*, Chin. Opt. Lett. **23**(6), 061901 (2025).
5. J. Lin, F. Bo, Y. Cheng, *et al.*, Photon. Res. **8** (12), 1910–1936 (2020).
6. B. Fu, R. Gao, N. Yao, *et al.*, Opto-Electron. Adv. **7**(8), 240061 (2024).
7. Z. Cai, Y. Liu, Z. Liu, *et al.*, Optica **13**(8), 1460–1468 (2026).
8. C. Li, N. Yao, H. Yu, *et al.*, Phys. Rev. Lett. **134**(21), 213801 (2025).
9. S. Liu, Y. Zheng, and X. Chen, Opt. Lett. **42**(18), 3626–3629 (2017).
10. L. Zhang, Z. Hao, X. Wu, *et al.*, Laser Photonics Rev. **20**(9), e01999 (2025).
11. J. Lin, N. Yao, Z. Hao, J. Zhang, *et al.*, Phys. Rev. Lett. **122**(17), 173903 (2019).
12. K. J. Vahala, Optical microcavities, Nature **424**(6950), 839–846 (2003).
13. I. Breunig, Laser Photonics Rev. **10**(4), 569–587 (2016).
14. C. Li, J. Deng, X. Huang, *et al.*, Phys. Rev. Lett. **136**(9), 093801 (2026).
15. Z.-Y. Wang, X. Wu, X. Xiong, *et al.*, Sci. Adv. **11**(18), eadu7605 (2025).
16. T. J. Kippenberg, A. L. Gaeta, M. Lipson, *et al.*, Science **361**(6402), eaan8083 (2018).
17. J. Leuthold, C. Koos, and W. Freude, Nat. Photonics **4**(8), 535–544 (2010).
18. Y. Zhu, B. Fu, Z, Fang, *et al.*, Opto-Electron. Adv. **9**(2), 250274 (2026).
19. X. Li, Y. Xie, S. Ding, *et al.*, Sci. Adv. **12**(1), aeb3400 (2026).
20. M. Li, G. Zhao, J. Lin, *et al.*, Opt. Express **32**(17), 29566–29572 (2024).
21. J. Liu, Q. Hou, Y. Zhou, *et al.*, Chin. Opt. Lett. **23**, 071302 (2025).
22. X. Zheng, Y. Yang, R. Gao, *et al.*, Opt. Lett. **51**(19), 5437–5440 (2026).
23. Z. Liu, W. Yao, M. You, *et al.*, ACS Photon. **12**, 684–695 (2025).
24. H. Jung, S. Yu, D. Carlson, *et al.*, Optica **8**(6), 811–817 (2021).
25. Q. Zhao, R. O. Behunin, P. T. Rakich, *et al.*, APL Photonics **5**(11), 116103 (2020).
26. G. M. Brodnik, G. Spektor, L. M. Williams, *et al.*, Nature 652(8111), 899–905 (2026).
27. Z. Geng, W. Cheng, Z. Yan, et al., Opt. Express **32**(7), 12291–12302 (2024).
28. X. Zhang, S. Guo, R. Yi, *et al.*, Optica **13**(10), 2135–2143 (2026).